# ROSAT/HRI and ASCA observations of the most luminous X-ray cluster RXJ1347.5-1145

S. Schindler[1,2,*], M. Hattori[1,3], D.M. Neumann[1], H. Böhringer[1]

[1] Max-Planck-Institut für extraterrestrische Physik, Giessenbachstraße, D-85478 Garching, Germany
[2] Max-Planck-Institut für Astrophysik, Karl-Schwarzschild-Straße 1, D-85478 Garching, Germany
[3] Astronomical Institute, Tôhoku University, Aoba Aramaki, Sendai 980, Japan

**Abstract.** We report on the X-ray properties of the exceptional X-ray cluster RXJ1347.5-1145 at $z$=0.451. We confirm that it is with a luminosity $L_X(bol) = 2 \times 10^{46}$ erg/s the most luminous X-ray cluster discovered to date. The mass of the cluster within 1.7 Mpc is $9.8 \times 10^{14} \mathcal{M}_\odot$. A comparison of the central X-ray mass and the mass determined from a simple gravitational lens model shows a discrepancy of a factor of 2-3 with the X-ray mass being smaller. The temperature of the cluster is $9.3^{+1.1}_{-1.0}$ keV. We detect a strong FeK line corresponding to a metallicity of $0.33 \pm 0.10$ in solar units, which is an unexpectedly high value for a distant and hot cluster. There are several hints that the cluster contains an extremely strong cooling flow. With the usual assumptions we derive formally a mass accretion rate of more than 3000 $\mathcal{M}_\odot$/yr indicating that this may be the largest cooling flow detected so far. To find these extreme properties in this distant cluster which can be taken as an indication of a well relaxed and old system is of high importance for the theory of formation and evolution of clusters.

## 1. Introduction

The cluster RXJ1347.5-1145 found in the ROSAT All Sky Survey is exceptional in many aspects. It is the most X-ray luminous cluster discovered to date (Schindler et al. 1995). Therefore this distant cluster (z=0.451) is an ideal target to analyse the X-ray properties of distant clusters. Of special interest for the formation and evolution of galaxy clusters are the cluster mass and the composition of the intracluster gas. Furthermore, the time evolution of cooling flows can be addressed.

Clusters can be used to put constraints to cosmological models. The cluster X-ray temperature function, the cluster X-ray luminosity function and the cluster mass function are providing information about cosmological parameters e.g. the shape and the amplitude of the power spectrum of primordial density fluctuations (see e.g. Edge et al. 1990; Henry & Arnaud 1991; Henry et al. 1992; Oukbir & Blanchard 1992; Bahcall & Cen 1993; Böhringer & Schindler 1994).

The masses of clusters can be determined in different ways: e.g. by using the X-ray emitting gas as a tracer for the cluster potential or by the gravitational lens effect. Previously, discrepancies were found in some clusters between the X-ray mass distribution and the mass distribution derived by strong lensing. In A2218 and A1689 the mass from strong lensing was found to be about 2 times higher than the X-ray mass (Miralda-Escudé & Babul 1995; Tyson & Fischer 1995). But there is also a counterexample: in PKS0745-191 Allen et al. (1995) found good agreement between strong lensing and X-ray mass with a multiphase cooling flow model. For many clusters very small core radii around 50 kpc were found by lens models: A370 and Cl0024+16 (Smail et al. 1995a), MS2137-23 (Mellier et al. 1993), A2218 (Kneib et al. 1995), A2219 (Smail et al. 1995c), Cl2236-04 (Kneib et al. 1994) while X-ray core radii for many clusters are around 250 kpc (Jones & Forman 1984). On the other hand, the masses of weak lensing and X-ray are in quite good agreement e.g. in A2218 (Squires et al. 1995), A2163 (Squires et al. 1996), Cl1455+22 and Cl0016+16 (Smail et al. 1995b).

RXJ1347.5-1145 is well suited for a comparison of lensing mass and X-ray mass, because it is very X-ray luminous and it was found to act as gravitational lens: two bright arcs were discovered (Schindler et al. 1995) in the course of the ESO Key Programme "Redshift Survey of ROSAT Clusters" (Böhringer 1994a; Guzzo et al. 1995). The arcs are located opposite to each other with respect to the cluster centre – an ideal configuration for a mass estimate. The distance of the arcs from the centre is 35 arcseconds (= 240 kpc) which encompasses an exceptional large volume for strong lensing.

The mass estimates from lensing and from a virial approach using the velocity dispersion of the galaxies will be reported in a forthcoming paper. In this paper we determine the X-ray mass by combining data from a ROSAT and an ASCA observation.

[*] e-mail: sas@mpa-garching.mpg.de

X-ray images of many clusters show the emission to be strongly peaked in the central region with the central cooling times of these clusters being smaller than a Hubble time. These central regions form cooling flows (see Fabian 1994). It is still uncertain whether cooling flows were stronger or weaker in the past. There is some observational evidence that cooling flows are getting stronger with time. Henriksen (1993) concluded from the X-ray analysis of bimodal clusters that the formation of cooling flows may occur relatively late in the formation of a rich cluster. White (1988) finds that most cooling flows have accretion rates which are increasing with time. The analyses of distant clusters like RXJ1347.5-1145 can therefore help to solve the question of the evolution of cooling flows.

Clusters of galaxies have metallicities typically in the range between 0.2 and 0.5 in solar units (Arnaud et al. 1992; Yamashita 1994; Tsuru et al. 1996). This implies that the intracluster gas contains at least a significant fraction of processed gas, which has at some point been ejected from stars. The amount of metals in clusters and the dependence of the metal content with time is crucial for the question of the origin of the gas. The different mechanisms for the ejection of the gas into the intracluster gas – supernova explosions and stellar winds after a starburst phase, collisions of galaxies, ram pressure stripping by the intracluster gas, or ejection during the formation of the galaxies – yield a different time dependence. Because the current supernova rate is much too small to produce the required amount of gas it must have been much higher in an earlier phase (Arnaud et al. 1992; Renzini et al. 1993). If the gas is ejected from the forming galaxies an early metal enrichment is expected, too. On the other hand ram pressure stripping is not effective as long as there is not enough intracluster gas. As soon as the gas density increases to a value high enough – e.g. by collisions of galaxies – ram pressure stripping starts and increases the density even more, thus leading to a very rapid stripping of the gas.

After the description of the data in Sect. 2 we perform a spatial analysis of the ROSAT/HRI data in Sect. 3 and a spectral analysis of the ASCA data in Sect. 4. In Sects. 5 – 7 we use both data sets to investigate the luminosity, the cooling flow and the mass of RXJ1347.5-1145. The results are discussed in Sect. 8.

Throughout this paper we use $H_0 = 50 \text{km s}^{-1}\text{Mpc}^{-1}$, $\Omega_0=1$ and $q_0=0.5$.

## 2. Data

For the X-ray analysis of RXJ1347.5-1145 we use ROSAT (Trümper 1983) and ASCA (Tanaka et al. 1994) data of three different types of instruments, ROSAT/HRI, ASCA/GIS and ASCA/SIS. These three instruments are complementary: while the ROSAT/HRI provides a spatial resolution of 5 arcseconds FWHM without any spectral resolution in the energy band of 0.1-2.4 keV, ASCA/GIS provides a spectral resolution of 0.08 to 0.105 FWHM (at 6 keV) in the energy band of 0.7-10. keV. ASCA/SIS has a spectral resolution of 0.02 FWHM at 6 keV.

RXJ1347.5-1145 was observed with the ROSAT/HRI on January 28[th] and 29[th], 1995 with a total exposure time of 15760 seconds and with ASCA on January 17[th] and 18[th], 1995 with an exposure time of 58300 seconds.

The data from both ASCA instruments – GIS and SIS – are reduced by standard techniques. The SIS observation was carried out in 4CCD bright mode during high bit rate and 2CCD bright mode during medium bit rate. In the present study, we only use the high bit rate 4CCD bright mode data. Since most of the cluster emission is located on a single CCD in each SIS (chip 1 in SIS0 and chip 3 in SIS1), we restrict our analysis to data from these chips.

## 3. Spatial analysis

For the spatial analysis we use the ROSAT/HRI observation. The HRI image (Fig. 1) shows a strongly peaked emission and a slightly elongated structure in the outer parts. The countrate in the maximum is 0.55 counts/s/arcmin$^2$. The maximum X-ray emission is at $\alpha = 13^h 47^m 31^s$, $\delta = -11°45'11''$ (J2000) which is 3 arcseconds southwest of the position of the central cluster galaxy. This offset is smaller than the nominal pointing uncertainty of the HRI of about 6 arcseconds. Unfortunately, there is no other source in the pointing that could be used to check whether this offset is real or due to the pointing uncertainty. For the overlay in Fig. 2 we assume that the positions of the X-ray maximum and the central galaxy coincide.

As the X-ray emission of the cluster is almost spherically symmetric we derive from the HRI data a radial profile of the surface brightness (Fig. 3) using the X-ray maximum as the centre. We fit a $\beta$-model to the surface brightness (following Cavaliere & Fusco-Femiano 1976; Jones & Forman 1984)

$$\Sigma(r) = \Sigma_0 \left(1 + \left(\frac{r}{r_c}\right)^2\right)^{-3\beta+1/2}, \quad (1)$$

where $\Sigma_0$ is the central surface brightness, $r_c$ is the core radius, and $\beta$ is the slope parameter. We find the following parameters: $\Sigma_0 = 0.60 \pm 0.02$ counts/s/arcmin$^2$, $r_c = 8.4 \pm 1.8$ arcsec ($57 \pm 12$kpc), and $\beta = 0.56 \pm 0.04$ ($1\sigma$ errors). The results for the fit parameters do not change significantly when varying the radial binning or the outer cut-off radius. The extremely small value for the core radius might be slightly influenced by the fact that the emission cannot be traced very far out for such a distant cluster because of the relatively high background of the HRI so that we see mainly the centrally peaked emission. To show that the emission is really extended and not a point source emission we plot also the HRI point spread function into Fig. 3. The peaked emission is a strong indication for a cooling flow in this cluster.

Taking the eccentricity into account we fit the data also with an elliptical $\beta$-model with different semi-axes but only one $\beta$ value (Neumann 1996). The best fit $\beta$ and core radii for the major and minor axes are in agreement with the $\beta$ and the core radius (within the errors) found in the 1D fit. The eccentricity is found to be 0.24 and the position angle is -17° (counterclockwise, north is 0). For comparison we fit ellipses to different isophote levels (see Bender & Möllenhoff 1987). These fits yield position angles between -10° and -25° (counterclockwise, north is 0) in good agreement with the elliptical $\beta$ model. At a count rate level of $9 \times 10^{-2}$ counts/s/arcmin$^2$ ($\approx$ the third lowest contour in Fig.2) it has the maximum eccentricity of 0.4.

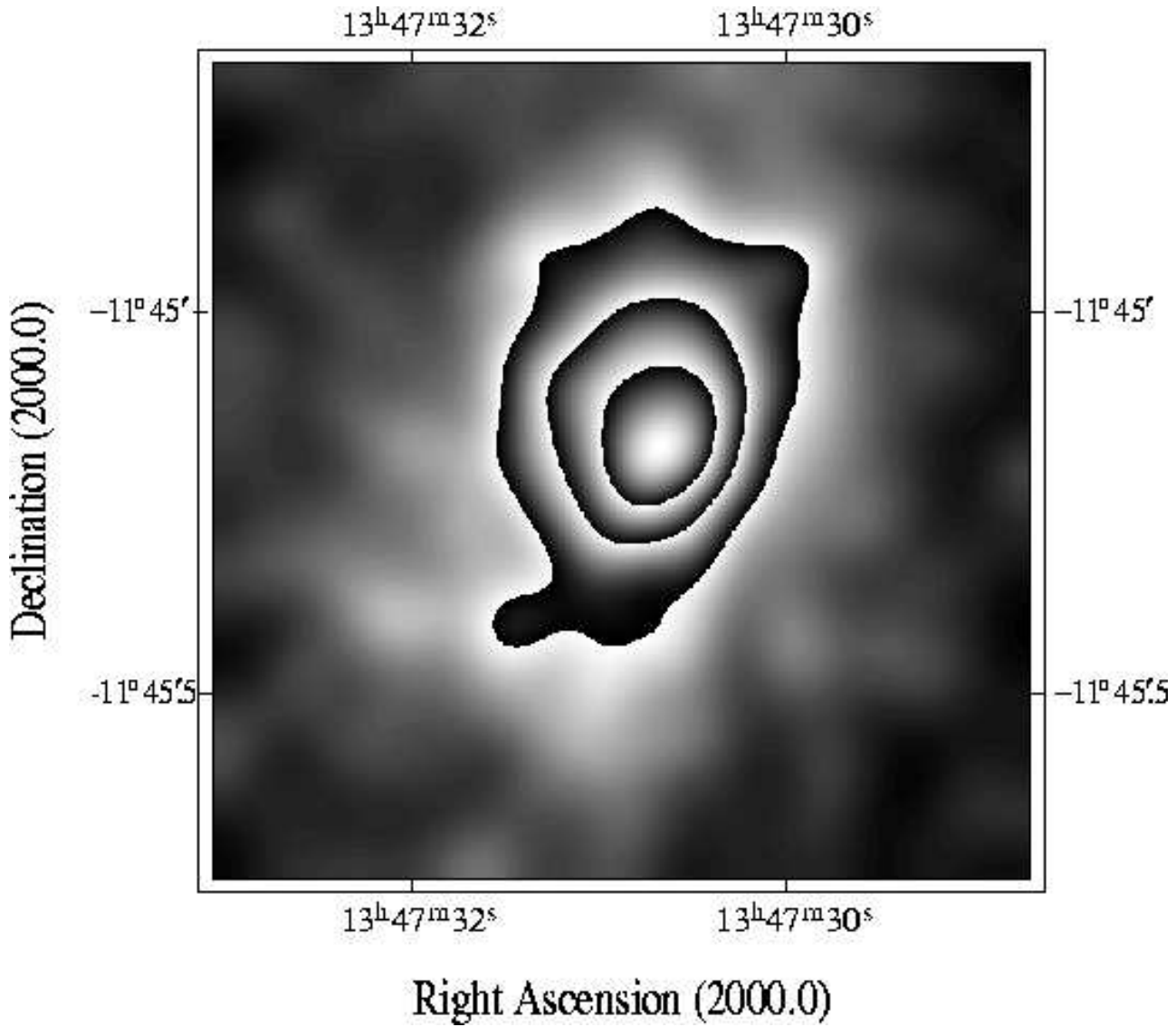

**Fig. 1.** ROSAT/HRI image of the cluster RXJ1347.5-1145. It is smoothed with a Gaussian filter of $\sigma = 2.5$ arcsec.

For further comparison we subtract a spherically symmetric image with the 1D fit parameters from an image with 0.5 arcseconds binning and $\sigma = 5$ arcseconds Gaussian smoothing. This procedure yields a $3\sigma$ excess emission in a region 10-15 arcseconds north of the maximum and a $4\sigma$ deficit of emission about 10 arcseconds south-west of the maximum. Another elongated region with an emission excess of 2-3$\sigma$ is about 20 arcseconds south-east of the maximum. This region is clearly visible in the second to forth lowest contours of Fig. 2.

## 4. Spectral analysis

We use the spectral-fitting package XSPEC to analyse the ASCA data. The data of both the GIS and SIS detectors are rebinned at a minimum signal-to-noise ratio of five in each of the spectral bins. The background is taken from blank sky fields at the same positions in the detector as the source. As there are faint contaminating sources in the background files for the GISs, we check our results with a background file in which the contaminating sources are masked (Ikebe et al. 1995) and find very similar fit parameters.

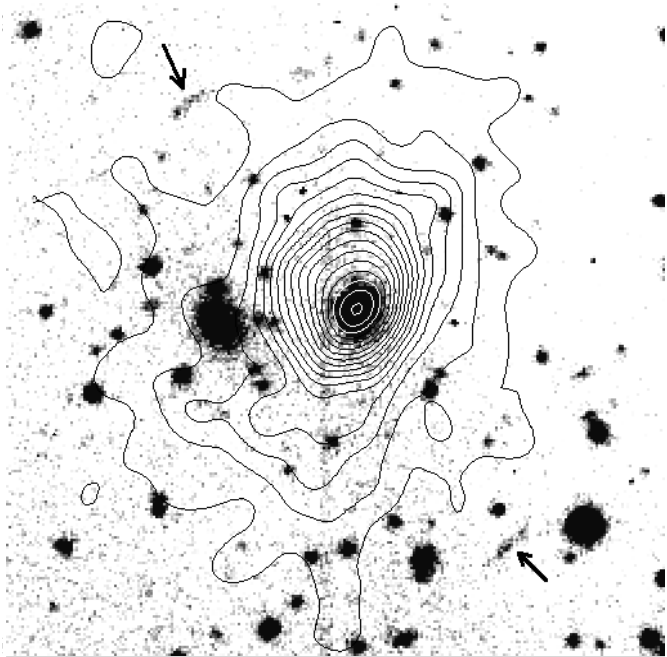

**Fig. 2.** X-ray contours of the HRI data superposed on an R image. The two images are aligned in such a way that the positions of the X-ray maximum and the central galaxy correspond. The X-ray image is smoothed with a Gaussian filter of $\sigma = 2.5$ arcsec. The contours are linearly spaced with $\Delta$countrate = 0.032 counts/s/arcmin$^2$ the highest contour line corresponding to 0.54 counts/s/arcmin$^2$. The positions of the arcs are marked by arrows. The size of the image is $1.4 \times 1.4$ arcmin$^2$ (North is up, East is left).

### 4.1. Temperature

To determine the temperature of the intracluster gas of RXJ1347.5-1145 we use the data of both GIS detectors on board of ASCA. As the point spread function of the ASCA X-ray telescope is energy dependent (Ikebe 1995) we determine at first an overall temperature of the cluster. Particularly, for such a strongly peaked surface brightness distribution it would be extremely difficult to derive a reliable temperature profile. To make sure that no high energy photons are lost due to the energy dependent point spread function we use the photons within relatively large radii: 5.9, 6.4, and 7.4 arcminutes. Only the photons with energies between 0.7 and 9 keV are taken into account. A typical spectrum is shown in Fig. 4.

We fit the spectra with a Raymond-Smith model (Raymond & Smith 1977). A selection of results is summarised in Table 1. A fit curve and the residuals are shown in Fig. 4. The fits yield a temperature of $9.3^{+1.1}_{-1.0}$ keV (90% confidence levels). No systematic differences are found in the temperature when fixing the redshift to the optically determined value $z=0.451$ (Schindler et al. 1995) or when leaving it free for the fit. When leaving the redshift free we find a value of $z = 0.45 \pm 0.02$ in agreement with the optically determined value. Fits with a Mewe-Kaastra model (Mewe et al. 1985, 1986; Kaastra 1992) give very similar results.

### 4.2. Hydrogen column density

Simultaneously with the temperature we fit the hydrogen column density. We find a value of $n_H = 0.10 \pm 0.04 \times 10^{22}$ cm$^{-2}$ (90% confidence levels). This value is about twice the Galactic value $n_H = 0.492 \times 10^{21}$ cm$^{-2}$ (Dickey & Lockman 1990). For most of the fits we find a deviation from the Galactic value of $2 - 3\sigma$. We cannot find any radial dependence of the $n_H$.

This enhanced absorption can be explained in two ways. Either there is an $n_H$ cloud right in front of the cluster which is not resolved on the $n_H$ map by Dickey & Lockman (1990). Or it could be a marginal indication of the presence of a cooling flow in the cluster which has produced cool gas in the cluster centre absorbing part of the X-ray emission.

Assuming that this excess absorption is intrinsic absorption in the cluster we fit the data with a model that takes intrinsic (=redshifted) absorption into account. We find in addition to the Galactic $n_H = 0.492 \times 10^{21}$ cm$^{-2}$ a hydrogen column density at the redshift of the cluster of $1.2 \pm 0.9 \times 10^{21}$ cm$^{-2}$. The values for the other fit parameters do not change when using the model with intrinsic absorption. If one takes into account that the cooling flow covers the cluster only partially this value could be even higher in the centre. The total mass of this cool gas would be of the order of $10^{12} \mathcal{M}_\odot$ within the cooling flow radius 200 kpc (see Sect. 6.).

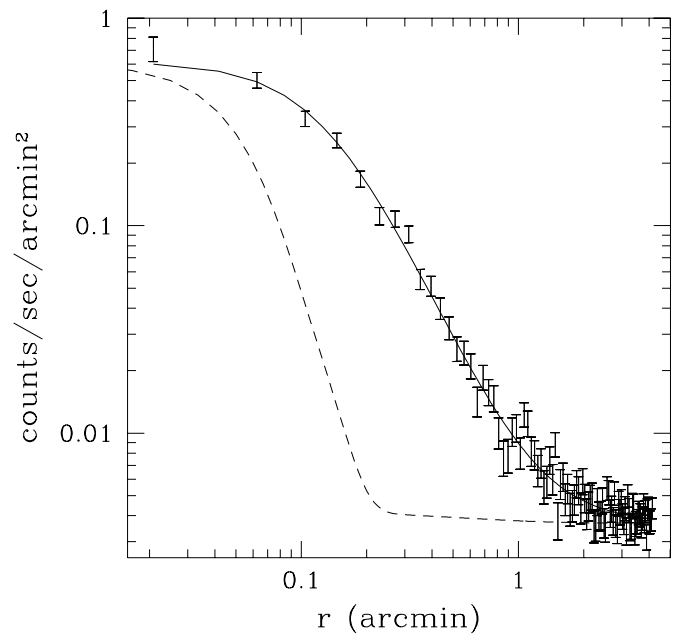

**Fig. 3.** Radial profile of the X-ray emission of RXJ1347.5-1145. The solid line is a fit with a $\beta$-model. Because of the strongly peaked emission we find a very small core radius of $r_c = 57 \pm 12$ kpc. To show that the cluster is extended we plot also the point spread function of the HRI (dashed curve) normalised to the central value of the $\beta$-model.

**Table 1.** Fit parameters of a selection of GIS spectra. The spectra of both GIS detectors are summed for the fit. The errors indicate 90% confidence limits.

| radius (arcmin) | temperature (keV) | metallicity (solar) | $n_H$ ($10^{22}$cm$^{-2}$) | $n_H$(redshifted) ($10^{22}$cm$^{-2}$) | redshift | reduced $\chi^2$ | DOF |
|---|---|---|---|---|---|---|---|
| 7.4 | $9.2^{+0.9}_{-0.8}$ | $0.32^{+0.10}_{-0.09}$ | $0.10^{+0.04}_{-0.04}$ | - | $0.44^{+0.03}_{-0.01}$ | 0.94 | 377 |
| 7.4 | $9.3^{+0.9}_{-0.8}$ | $0.32^{+0.10}_{-0.09}$ | $0.10^{+0.04}_{-0.04}$ | - | 0.451(fixed) | 0.94 | 378 |
| 7.4 | $9.3^{+1.0}_{-0.8}$ | $0.32^{+0.10}_{-0.09}$ | 0.0492(fixed) | $0.13 \pm 0.08$ | 0.451(fixed) | 0.94 | 378 |
| 6.4 | $9.1^{+0.9}_{-0.8}$ | $0.32^{+0.10}_{-0.08}$ | $0.10^{+0.04}_{-0.03}$ | - | 0.451(fixed) | 0.96 | 365 |
| 5.9 | $9.9^{+1.1}_{-0.9}$ | $0.34^{+0.10}_{-0.1}$ | $0.09^{+0.04}_{-0.03}$ | - | 0.451(fixed) | 0.94 | 359 |
| 3.0 | $11.8^{+1.6}_{-1.0}$ | $0.39^{+0.12}_{-0.13}$ | $0.06^{+0.03}_{-0.03}$ | - | 0.451(fixed) | 1.01 | 335 |
| 1.5 | $8.5^{+1.0}_{-0.9}$ | $0.41^{+0.13}_{-0.12}$ | $0.11^{+0.05}_{-0.04}$ | - | 0.451(fixed) | 0.94 | 214 |

*4.3. Metallicity*

For the spectral fitting we use the relative metal abundances according to Anders & Grevesse (1989). In the spectrum (Fig. 4) a very prominent Fe K line around 5 keV (redshifted) is visible. By fitting the spectrum with a Raymond-Smith model we find a metallicity of $0.33 \pm 0.10$ (90% confidence levels) in solar units. This value is determined for the overall cluster with radii of 5.9, 6.4, and 7.4 arcminutes. When going to a smaller radius of 1.5 arcminutes one finds a small increase of the metal abundances $m = 0.41^{+0.13}_{-0.12}$. This increase cannot be an artifact of the energy-dependent point spread function as the scattering of the high energy photons would decrease the intensity of the iron line. As the strength of the iron line is only weakly dependent of the temperature in the range of 1-10 keV, the metallicity is not expected to decrease even if there is a cool component in the gas.

**5. Luminosity**

In a previous analysis (Schindler et al. 1995) the X-ray luminosity could be inferred only from the 118 source counts detected in the ROSAT All Sky Survey. The pointed observations in two different energy bands now provide more accurate numbers.

In the ROSAT/HRI profile (Fig. 3) the X-ray emission can be traced out to a radius of 4.2 arcminutes. Within this radius we find 2200 source counts corresponding to a luminosity $7.3 \pm 0.8 \times 10^{45}$ erg/s in the ROSAT band (0.1 – 2.4 keV) and to a bolometric luminosity of $2.1 \pm 0.4 \times 10^{46}$ erg/s for $T$=9.3 keV and $n_H = 1.0 \times 10^{21}$cm$^{-2}$. The error includes the errors in $T$ and $n_H$.

From the ASCA/GIS we find a net count rate of 0.18 counts/s in 0.7-9 keV within a radius of 6.4 arcminutes corresponding to a 2-10 keV restframe luminosity of $7.58 \pm 0.07 \times 10^{45}$ erg/s for the above values of temperature, metallicity and hydrogen column density. This luminosity corresponds to a bolometric luminosity of $1.7 \times 10^{46}$ erg/s. The difference of the bolometric luminosities from ASCA/GIS and HRI could be due to a central cool component (see Sect 6.).

**6. Cooling flow analysis**

Using the fit parameters from the $\beta$-model $n_{e0} = 0.094$ cm$^{-3}$, $r_c = 57$kpc, $\beta = 0.56$ and the temperature from the ASCA spectrum T=9.3 keV we find a central cooling time of $1.2 \times 10^9$ yr. With the usual assumptions (e.g. cooling time smaller than $10^{10}$ yr) we derive a cooling flow radius of 29 arcseconds = 200 kpc. The fraction of luminosity emitted from the cooling flow region ($r \leq 200$ kpc) can be estimated to be 43% of the total cluster emission. With this fraction and the equation

$$L_{\text{cool}} = \frac{5}{2} \frac{\dot{M}}{\mu m} kT, \qquad (2)$$

we find a for temperature of 9.3 keV a huge mass accretion rate of 3900 $\mathcal{M}_\odot$/yr. This is an upper limit for the accretion rate because the gain of potential energy of the gas flowing in are not taken into account. With a more realistic cooling flow analysis assuming that the condensed matter supplies the gas with its complete loss of thermal energy (see Arnaud 1988; Neumann & Böhringer 1995) and the same fit parameters we find a mass accretion rate around 3500 $\mathcal{M}_\odot$/yr.

These numbers derived with the usual assumptions are the ones to be used for comparison with other mass accretion rates. Smaller numbers are obtained when we take into account that the cluster is in fact younger. For our cosmological parameters its age is only $7.4 \times 10^9$ yr. Assuming that the formation process lasted $3 \times 10^9$ yr and the cooling time is $4.4 \times 10^9$ yr we find a smaller cooling flow radius of 110 kpc and consequently also a smaller mass accretion rate of 2300 $\mathcal{M}_\odot$/yr.

To see whether there is any signature of a cool component in the GIS data we analyse the central region within a radius of 1.5 arcminutes. We find a temperature of $8.5^{+1.0}_{-0.9}$keV (90% confidence level) – slightly smaller than the overall temperature. This could be either a signature of a cooling flow or, as well, it could be an artifact of the energy dependent point spread function of the ASCA X-ray Telescope (Ikebe 1995). The fit gives about the same result for the hydrogen column density as the fit to the overall data. A fit to data within an intermediate radius (3 arcminutes) gives a temperature of $11.8^{+1.6}_{-1.0}$ keV i.e. a value that is higher than the overall temperature as well as the central temperature. These results show no clear indication of a cooling flow, but seem to be a hint for a complex temperature structure in the cluster.

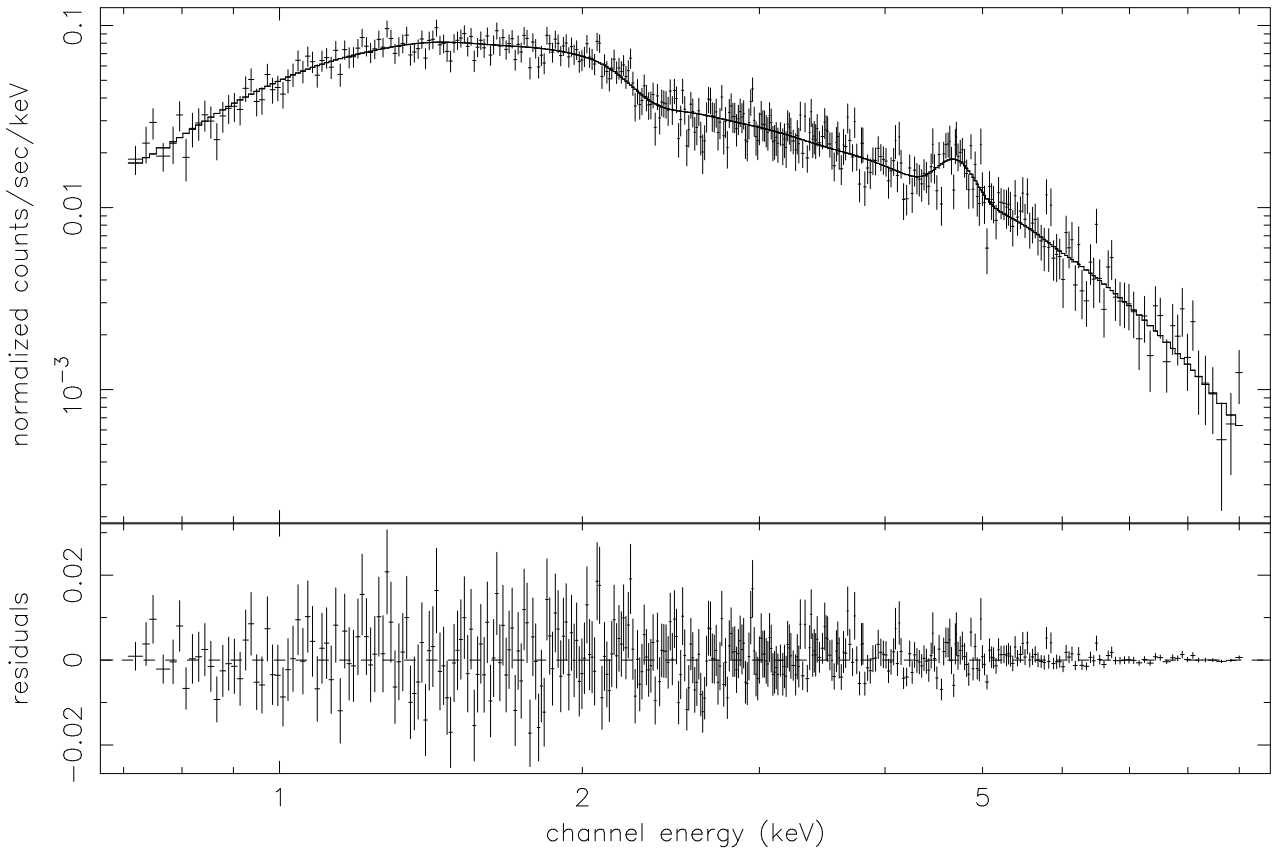

**Fig. 4.** Spectrum of both GIS detectors within a 6.4 arcminute radius. Around 5 keV the (redshifted) FeK line is visible, which corresponds to a metallicity of 0.33 in solar units. A fit with a Raymond-Smith model (solid line) and the residuals are shown, too.

Fitting the central data (within 1.5 arcminutes and within 3 arcminutes) with a two temperature fit does not show any evidence for a cool component: the $\chi^2$ does not improve. The normalisation of the low temperature component is not well determined and always compatible with zero, i.e. from these data the existence of cool gas can neither be confirmed nor ruled out.

However, a hint for a cool component comes from a comparison of the ASCA/GIS and the ROSAT count rate: estimating from the ASCA/GIS count rate of 0.18 counts/s within $r = 6.4$ arcminutes with the above mentioned fit parameters for the overall cluster we find an HRI countrate of 0.11 counts/s but 0.14 counts/s are observed. The larger HRI count rate could be explained by a cool component emitting more counts in the softer energy band of the HRI.

The ASCA/SIS with its slightly softer energy range and better spectral and spatial resolution compared to the GIS is better suited for the analysis of a cool component. We use the bright mode high bit rate data, which have an effective exposure time of $4.5 \times 10^4$ seconds, and integrate the cluster emission in apertures of 0.5, 1.0, 1.5 and 4.3 arcminutes radius. With the 4.3 arcminutes data we correct for the gain shift (Dotani et al. 1995) with the FeK line. Compared to the optical value the FeK line blueshifted by 3%. As we cannot correct for this shift in the background data and the response files we can perform only a semi-quantitative analysis. Although the absolute values obtained by the SIS spectra can have systematic values it is worthwhile to analyse to data because of their good energy resolution.

The SIS spectrum within a radius of 1.5 arcminutes is shown in Fig. 5. In the spectrum the FeK $\alpha$ and FeK $\beta$ lines are distinguishable. Unfortunately, the statistics of the spectrum is too low to determine a line ratio. The existence of a strong FeK $\beta$ line is evidence for the presence of gas at a temperature larger than 6 keV. We see no signature of an FeL line (for this redshift it should be at 0.7 keV) which would be a typical indication of cool gas (see e.g. Fabian et al. 1994a). For the radii between 0.5 and 1.5 arcminutes we do the same analysis as for the GIS data with one and two temperature fits. We find the same behaviour: the $\chi^2$ does not improve when going from a one to a two temperature fit. We obtain the same result whether we fix the fit parameters – the hot temperature, the cool temperature and the metallicity – or not. The normalisation of the cool component is not well determined and always compatible with zero.

To see whether the existence of the strong cooling flow indicated by the HRI data is consistent with ASCA data we fix the ratio between the normalisations of the hot and the cool component. We use again the information from the HRI profile, that 43% of the total emission in the ROSAT band is emitted from the cooling flow region ($r \leq 200$ kpc). This fraction must be transformed to the ASCA energy range where the emission ratio between a 1 and a 10 keV gas is about 1:2. We fit the SIS data with this fixed ratio. The result is acceptable but there is no improvement of $\chi^2$ implying that there is no significant evidence for a cool component but the existence of a strong cooling flow indicated by the HRI data can not be ruled out.

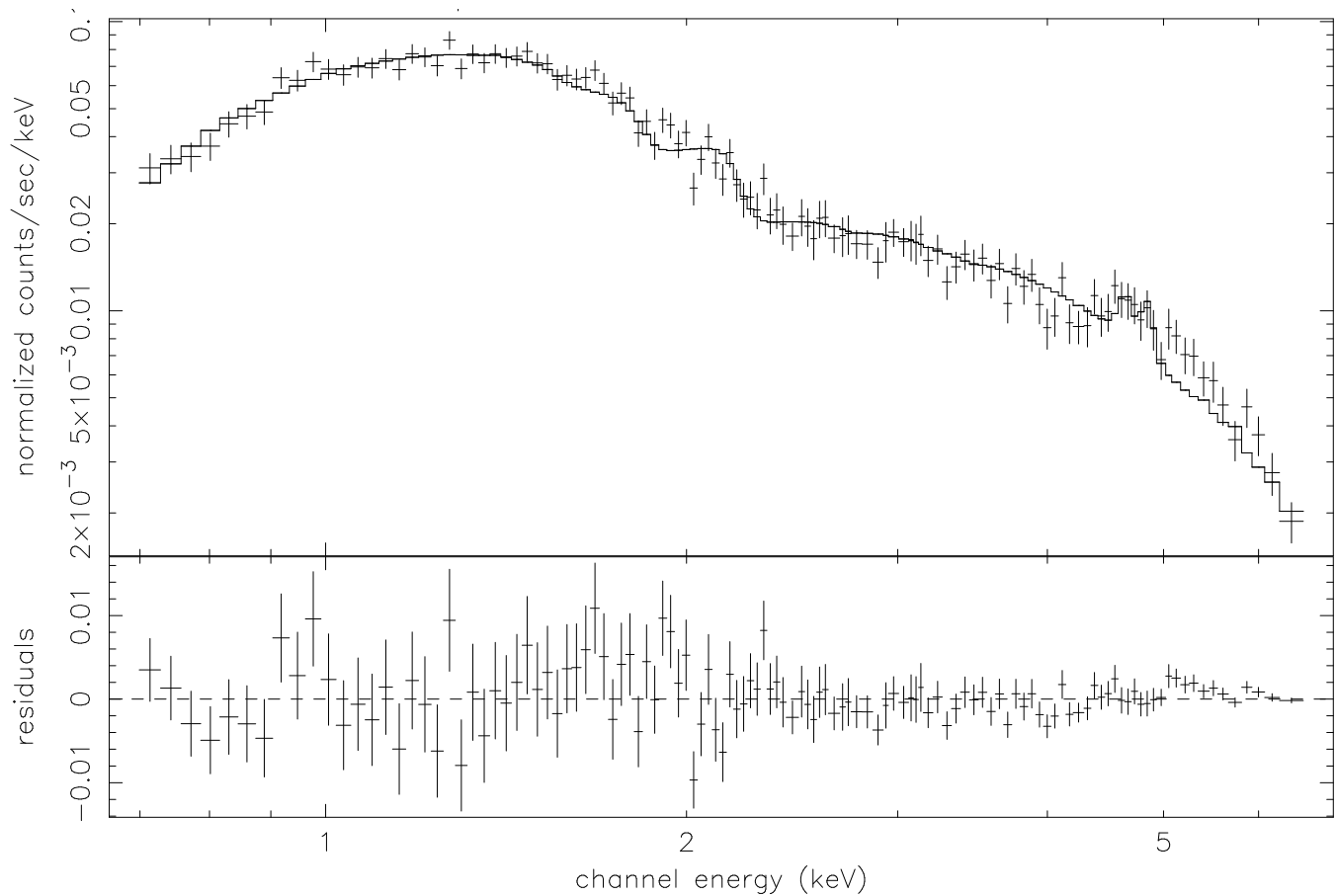

**Fig. 5.** Spectrum of both SIS detectors of bright mode medium and high bit rate within a 1.5 arcminute radius. The FeK $\alpha$ and $\beta$ lines around 5 keV are clearly visible. The solid line shows a Raymond-Smith fit with a metallicity of 0.34 in solar units.

## 7. Mass determination

The parameters of the $\beta$-model can be used to make a deprojection of the 2D image to get the three dimensional density distribution. The profile of the integrated gas mass is shown in Fig. 6. Within a radius of 1 Mpc the gas mass amounts to $2.0 \times 10^{14} \mathcal{M}_\odot$. Extrapolated to a radius of 3 Mpc we find $8.9 \times 10^{14} \mathcal{M}_\odot$. As the emissivity of the gas in the ROSAT energy band is almost independent of the temperature (within the temperature range of 2-10 keV it changes only by 6%), we can derive the gas density distribution hardly affected by the uncertainty of the temperature estimate. The only uncertainty are local unresolved inhomogeneities of the gas which would result in an overestimation of the true gas mass.

With the additional assumption of hydrostatic equilibrium, the integrated total mass can be calculated from the equation

$$M(r) = \frac{-kr}{\mu m_p G} T \left( \frac{d \ln \rho}{d \ln r} + \frac{d \ln T}{d \ln r} \right), \quad (3)$$

where $\rho$ and $T$ are the density and the temperature of the intracluster gas, and $r$, $k$, $\mu$, $m_p$, and $G$ are the radius, the Boltzmann constant, the molecular weight, the proton mass, and the gravitational constant, respectively.

As discussed in the previous section no clear evidence for a temperature drop to the centre could be found. Therefore we use a constant temperature of 9.3 keV for the mass determination. Fig. 6 shows the profile of the integrated mass. Within 1 Mpc we find a total mass of $5.8 \times 10^{14} \mathcal{M}_\odot$ and within the radius of the arcs (240 kpc) $1.3 \times 10^{14} \mathcal{M}_\odot$. Extrapolated to a radius of 3 Mpc the total mass is $1.7 \times 10^{15} \mathcal{M}_\odot$. The dashed lines in Fig. 6 show the errors which come from the uncertainty in the overall temperature. But these errors are probably much smaller than the errors introduced by the assumption of isothermality in particular for the central region. The errors coming from the uncertainty in the density profile are comparably small.

We estimate the error in the total mass due to the assumptions of hydrostatic equilibrium and spherical symmetry to be about $\pm 15 - 20\%$ in the outer parts from comparison with numerical simulations (Schindler 1996). In the centre they can be considerably higher. We estimate that they can amount to $\pm 50\%$ in the cooling flow region, because from the GIS data we find hints for a complex temperature structure, which cannot be taken into account with our isothermal model.

For a comparison with mass estimates determined from the gravitational lensing effect, we also calculate the total mass of the cluster, as seen inside a certain angle, basically integrating a spherically symmetric three dimensional mass distribution in cylindrical shells with cylinder axis parallel to the line of sight or integrating the cluster surface mass density $\Sigma(\theta)$ outward (see Fig. 6). At the radius of the arcs we find $2.1 \times 10^{14} \mathcal{M}_\odot$. This mass is considerably smaller than the mass expected from a simple point mass lensing model $4.4$-$7.8 \times 10^{14} \mathcal{M}_\odot$ (Schindler et al. 1995) with estimated redshifts of the sources of $z = 0.7 - 1.2$. For checking the possibility that we see with the HRI only the peak of the cooling flow emission and the overall cluster emission is hidden in the background, we calculate the surface mass density also with standard profile parameters: $r_c$ = 250 kpc and $\beta = 0.65$. With these parameters we find an even smaller mass of $1.4$-$1.9 \times 10^{14} \mathcal{M}_\odot$.

From the profiles of the gas mass and the total mass we can infer the gas-to-total-mass ratio. It increases with radius: at 240 kpc it is 19%, at 1 Mpc it is 34%. Extrapolated to 3 Mpc we find even 52%. The estimated errors are basically the same as for the total mass. As the ratio for 1 and 3 Mpc is unusually high it could be an indication that we underestimate the total mass.

## 8. Discussion and Conclusions

RXJ1347.5-1145 has many exceptional X-ray properties. It shows very high values for the luminosity, the metallicity and the cooling flow rate.

RXJ1347.5-1145 has an extremely high luminosity of $L_X(bol) = 2 \times 10^{46}$ erg/s. As this X-ray luminosity is the largest found to date in a cluster of galaxies the question arises whether it could be contaminated by another X-ray source e.g. a quasar. The optical spectrum of the central galaxy (Schindler et al. 1995) which is located almost at the maximum of the X-ray emission clearly rules out this possibility as well as the extent of the X-ray source in the HRI image. In the centre of the cluster could be an AGN but the X-ray emission of an AGN with narrow emission lines at that redshift would be comparably small.

A high redshift cluster with this high luminosity does not underline the evolution of the X-ray luminosity function found by Henry et al. (1992). They found a decreasing number of high-luminosity clusters when going from a redshift $z = 0.14$ to $z = 0.6$. When assuming that mergers of smaller subclusters increase the depth of the potential well and the amount of gas, and hence the luminosity, the finding of a cluster which has formed already relatively early points to a low merger rate in the recent past, i.e. to a low $\Omega$.

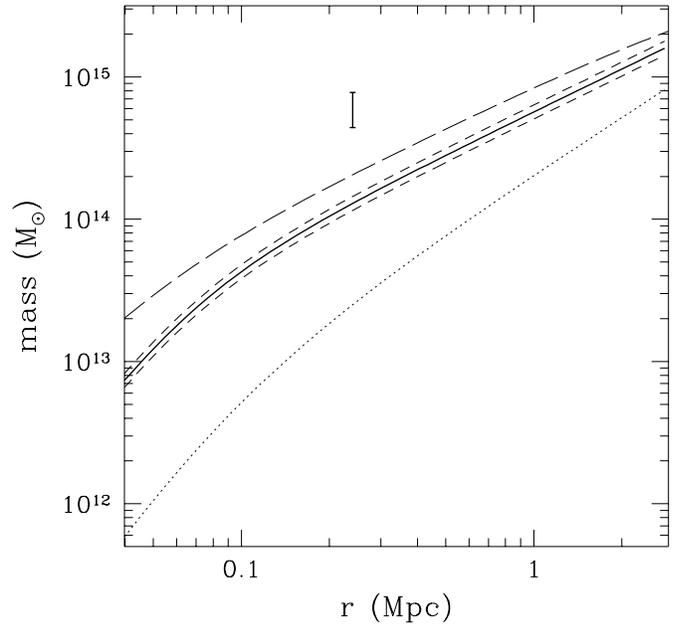

**Fig. 6.** Profile of the integrated total mass (full line) with errors coming from the uncertainty in the overall temperature (short-dashed lines). The dotted line shows the profile of the integrated gas mass. For comparison with mass determinations by the gravitational lens effect, the long-dashed line depicts the integrated surface mass density. The mass estimate by the gravitational lens effect is marked by the error bar.

**Table 2.** Summary of the X-ray properties of RXJ1347.5-1145

| | |
|---|---|
| $L_X(bol)$[HRI] | $21 \pm 4 \times 10^{45}$ erg/s |
| $L_X(0.1$-$2.4\text{keV})$[HRI] | $7.3 \pm 0.8 \times 10^{45}$ erg/s |
| $L_X(2$-$10\text{keV})$[GIS] | $7.6 \pm 0.1 \times 10^{45}$ erg/s |
| countrate(0.1-2.4keV)[HRI] | 0.14 counts/s |
| countrate(0.7-9.0keV)[GIS] | 0.18 counts/s |
| $r_c$ | $8.4 \pm 1.8$ arcsec ($57 \pm 12$ kpc) |
| $\beta$ | $0.56 \pm 0.04$ |
| $T$ | $9.3^{+1.1}_{-1.0}$ keV |
| metallicity | $0.33 \pm 0.10$ solar |
| $M_{gas}(r < 1\text{Mpc})$ | $2.0 \times 10^{14} \mathcal{M}_\odot$ |
| $M_{gas}(r < 3\text{Mpc})$ | $8.9 \times 10^{14} \mathcal{M}_\odot$ |
| $M_{tot}(r < 1\text{Mpc})$ | $5.8 \times 10^{14} \mathcal{M}_\odot$ |
| $M_{tot}(r < 3\text{Mpc})$ | $17 \times 10^{14} \mathcal{M}_\odot$ |
| gas mass fraction ($< 1$Mpc) | 34% |
| gas mass fraction ($< 3$Mpc) | 52% |
| cooling flow radius | 29 arcsec (200 kpc) |
| central cooling time | $1.2 \times 10^9$ yr |
| mass accretion rate | $\gtrsim 3000 \mathcal{M}_\odot$/yr |
| $n_H$ | $0.10 \pm 0.04 \times 10^{22}$ cm$^{-2}$ |
| redshift | 0.451 |

Although we cannot find clear evidence in the spectra for cool gas in the central region we find several other indications of a cooling flow. The surface brightness profile is very strongly peaked which is the typical signature of a cooling flow. Furthermore, there is excess absorption compared to the Galactic absorption. We find an $n_H$ about twice as high as the Galactic value. This high column density was found in a number of cooling flow clusters (White et al. 1991) and can be interpreted as cool gas which has been produced in the cooling flow. We see no hint for the absorption in the central part of the profile. Neither is there any excess emission visible on top of the King profile in the cooling flow region. Therefore, it might be that both effects compensate each other. Another indication for a cooling flow is that RXJ1347.5-1145 fits into the $L_X - T$-relation of the clusters with the strongest cooling flows (see Fabian et al. 1994b). The finding of emission lines in the spectrum of the central galaxy (Schindler et al. 1995) is also compatible with the existence of a cooling flow.

From a cooling flow analysis we derive formally a mass accretion rate of more than 3000 $\mathcal{M}_\odot$/yr within a radius of 200 kpc for a cooling time of $10^{10}$ yr. When comparing this value with other mass accretion rates derived with the same assumptions the cooling flow in RXJ1347.5-1145 is the strongest one found to date. The finding of a very strong cooling flow at high redshift does not point to a scenario which predicts a continuous increase of the mass accretion rate with time. Other strong cooling flows have been in found in clusters with a large range of redshifts: e.g. in A478: 900 $\mathcal{M}_\odot$/yr (White et al. 1994) at z=0.0881, in PKS0745-191: 1000 $\mathcal{M}_\odot$/yr (Allen et al. 1995) at z=0.1028, in A2390: 850 $\mathcal{M}_\odot$/yr (Pierre et al. 1995) at z=0.232, in Zwicky 3146: 1200 $\mathcal{M}_\odot$/yr (Edge et al. 1994) at z=0.2906, in P01904+4109: 1000 $\mathcal{M}_\odot$/yr (Fabian & Crawford 1995) at z=0.442. As there is no tendency for an increasing mass accretion rate with time a picture with the cooling flows being from time to time disrupted by mergers and reestablishing after a cooling time seems to be more probable.

When taking into account that RXJ1347.5-1145 probably did not have $10^{10}$ yr to cool we find smaller accretion rates, e.g. for a cooling time of $4.4 \times 10^9$ yr we find a cooling flow radius of 110 kpc and a mass accretion rate of 2300 $\mathcal{M}_\odot$/yr. Also, using a Hubble constant larger than $H_0 = 50$km s$^{-1}$Mpc$^{-1}$ would decrease the mass accretion rate.

The metallicity of the gas $m = 0.33 \pm 0.10$ in RXJ1347.5-1145 is exceptionally high for the redshift. The most distant cluster with a detected Fe line (corresponding to a metallicity of $m = 0.5$) used to be A370 at $z = 0.367$ (Bautz et al. 1994), but there is probably a contaminating source in this cluster (Böhringer, private communication). A newly found very distant cluster at $z \approx 1$ (Hattori et al. 1996) shows also a high metallicity. These findings favour the models by Arnaud et al. (1992) and Elbaz et al. (1995) with enhanced star formation at early epochs. But on the other hand there are also counterexamples: e.g. Cl0016+16 at $z = 0.55$ has a metal abundance compatible with zero: $m < 0.167$ (Tsuru et al. 1996). Obviously, there is a large scatter in the metal abundances of distant clusters, which makes the study of the metal enrichment mechanisms even more interesting.

Compared with the X-ray temperature - metallicity relation from EXOSAT and GINGA for nearby clusters (Arnaud et al. 1992) and from ASCA (Ohashi 1995; Tsuru 1996) RXJ1347.5-1145 is slightly on the high metallicity side but still within the errors. Given that the cluster could be even hotter if one could separate the cooling flow emission and the emission of the rest of the cluster, this cluster does not support the trend of a decreasing metallicity with increasing temperature. Its metallicity of $m = 0.33$ is a typical value for clusters with temperatures between 1 and 5 keV. Therefore the assumption that the decrease of metallicity with temperature is related to a decrease of metallicity with richness – where the gas ejected from galaxies is diluted by a larger amount of primordial gas – is not supported by this observation.

Because of the large and energy-dependent point spread function of the ASCA X-ray Telescope we cannot perform a spatially resolved analysis of the temperature distribution. But when integrating the photons within different radii we find variations of the temperature, which can be hints for complex temperature structures. Such complex structures occur during and after the merging of subclusters (see Schindler & Müller 1993). Also the elongation of the emission in the outer parts is an indication that the cluster is not completely relaxed. On the other hand the strong cooling flow points to a relatively long interval in which the cluster could have evolved without any disturbances. Therefore, the dynamical state of RXJ1347.5-1145 cannot be determined conclusively.

The mass of the cluster within a radius of 1 Mpc is $5.8 \times 10^{14} \mathcal{M}_\odot$. Extrapolated to 4 Mpc we find a total mass of $2.5 \times 10^{15} \mathcal{M}_\odot$. This value is comparable to the mass of the Coma cluster [(1.1 - 2.8)$\times 10^{15} \mathcal{M}_\odot$ within 4 Mpc (Böhringer 1994b)]. Obviously, the mass of this cluster is smaller than it could have been expected from the extremely high luminosity. The reason is probably the strong cooling flow which contributes a significant fraction ($\approx 43\%$) to the total emission.

At 1 Mpc the ratio of gas-to-total-mass is 34%. This value is unusually high for the small radius. As it is derived by neglecting any temperature variations this large ratio could indicate that we underestimate the total mass because of temperature gradients.

The integrated surface mass density of the central part of the cluster (240 kpc) is $2.1 \times 10^{14} \mathcal{M}_\odot$. This value is considerably smaller than the mass expected from the gravitational lens effect (4.4-7.8$\times 10^{14} \mathcal{M}_\odot$). There are several possible explanations for this discrepancy. The lens model is very simple. It assumes a point mass at the centre of the cluster and the redshifts of the arcs are only estimated. A better model with a new lensing mass estimate will be published in a forthcoming paper.

There are some uncertainties on the X-ray side, too. We assume a constant temperature which is very probably not a good assumption for the central part of the cluster because of the cooling flow. To obtain a temperature profile of the core region one has to correct for the energy dependence of the point spread function of the ASCA X-ray Telescope. There is considerable effort made to develop a tool for this important correction (Ikebe 1995) and applying it to RXJ1347.5-1145 would certainly give new insights into the temperature structure. There are also several physical explanations for the discrepancy. If the cluster has an ellipticity along the line-of-sight or there is substructure in the cluster the mass would be underestimated by the X-ray method (Schindler 1996). Another possible explanation is non-thermal pressure support which can be provided by strong turbulence and equipartition magnetic fields (Loeb & Mao 1994).

To get better estimates for the lensing mass (weak and strong) lensing and to derive also a virial mass RXJ1347.5-1145 is a target for further optical studies.

*Acknowledgements.* It is a pleasure to thank Y. Tanaka and A. Kritsuk for useful discussions and Y. Ikebe for the cleaned background files. We also thank the ROSAT team and the ASCA team for providing the data. S.S. gratefully acknowledges the hospitality of the Astronomical Institute of the University of Basel. S.S. and H.B. thank the Verbundforschung for financial support.